\documentstyle[aps,prl,amssymb,multicol,graphicx]{revtex}

\begin{document}
%\draft
%\preprint{}

\title{Intrinsic tunneling spectra of $\rm Bi_2(Sr_{2-x}La_x)CuO_6$}

\author{ A.Yurgens,$^{1}$ D. Winkler,$^{1,2}$ T. Claeson,$^{1}$ S. Ono,$^{3}$
and Yoichi Ando$^{3}$}

\address{$^1$Chalmers University of Technology, S-41296 G\"oteborg,
Sweden}

\address{$^2$IMEGO Institute, Aschebergsgatan 46, S41133, G\"oteborg,
Sweden} \date{\today }

\address{$^3$Electrical Physics Department, Central Research Institute of Electric Power Industry (CRIEPI), Komae, Tokyo 201-8511, Japan}

\date{\today}
\maketitle
\begin{abstract}
We have measured intrinsic-tunneling spectra of a single  CuO-layer La-doped \linebreak $\rm Bi_2Sr_{2-x}La_xCuO_{6+\delta}$  (Bi2201-La$_x$).  Despite a difference of a factor of three in the optimal superconducting critical temperatures for Bi2201-La$_{0.4}$ and Bi2212 (32 and 95 K, respectively) and different spectral energy scales, we find that the pseudogap vanishes at a similar characteristic temperature $T^*\approx 230-300$~K for both compounds.  We find also that in Bi2201-La$_x$, PG humps are seen as sharp peaks and, in fact, even dominate the intrinsic spectra.
\end{abstract}

\vspace{ 6pt}

\begin{multicols}{2} 
\narrowtext

%\newpage
%----------------------------------------------------------------------------------------------------
The existence of a pseudogap (PG) in the normal-state quasi-particle density of states (DOS) of high-temperature superconductors (HTS) was revealed in many experiments \cite{Timusk}.  The PG phenomenon attracts considerable attention because it is believed to be a key to the understanding of high-temperature superconductivity.   Many groups tried to find a common relation between PG and superconductivity, and several models have been advanced \cite{Timusk}.  Two powerful surface probes,  angle-resolved photoemission spectroscopy (ARPES)  and scanning tunneling spectroscopy, can directly access the DOS.  They showed that in $\rm Bi_2Sr_2CaCu_2O_{8+\delta}$ (Bi2212)  and  $\rm Bi_2Sr_2Cu_1O_{6+\delta}$ (Bi2201),  a PG in the normal state develops into the superconducting gap (SG) once phase coherence is established below the superconducting critical temperature $T_c$ \cite{ARPES,Renner}.  This favored the idea that PG is intimately  connected to superconductivity and that it could be the signature of precursor pairing \cite{precursor}.  

However, the $c$-axis inter-plane (intrinsic) tunneling \cite{Kleiner}, which allows access to the DOS from "inside" a single crystal, resolved separate spectral features related to superconductivity and the pseudogap.  Sharp SG peaks and shallow PG "humps"  coexist at low temperatures  and have different temperature and magnetic-field dependencies \cite{HgBr,Krasnov}  implying different origins of the PG and superconductivity.   Most experiments have, so far, been  restricted to Bi2212, and it is especially important to study a compound different from Bi2212 in order to make general conclusions.

Here we present for the first time intrinsic-tunneling spectra of single CuO-layer La-doped $\rm Bi_2Sr_{2-x}La_xCuO_{6+\delta}$  (Bi2201-La$_x$), having a much lower superconducting critical temperature ($T_c^{max}\approx 32$ vs. 95~K in Bi2212). It turns out that the characteristic pseudogap temperature, $T^*\approx 230-300$~K, is nearly the same for both compounds. We also find that PG humps are seen as sharp peaks that dominate the intrinsic spectra in Bi2201-La$_x$. This implies that the dips, which have been stressed \cite{Zasadzinski}, are much likely a result of the superposition of SG- and PG peaks in the present experiments.

We measured intrinsic tunneling spectra across mesas about 15-150 nm high and from  $\sim 4\times 4$ to $10\times 10$ $\mu $m$^2$ in area. Such mesas comprise  $N = 10-100$ intrinsic tunnel junctions. The mesas were lithographically patterned on the surfaces of Bi2201-La$_x$ single crystals \cite{Ando}.   Details of the sample fabrication can be found elsewhere \cite{we_APL}.    

Fig.~\ref{dIV1} shows current-voltage (I-V) characteristics of optimally doped Bi2201-La$_{0.4}$ mesas with T$_c$ $\approx $ 32~K.  At small currents (main panel), there are $N$ = 18 hysteretic branches, every one corresponding to sequential switching of more and more intrinsic tunnel junctions from the superconducting to the normal state.   When the bias current becomes larger than the superconducting critical current $I_{c}^{max}$  of the last ("strongest") junction in the mesa, the I-V characteristic becomes single-valued, see  Fig.~\ref{dIV1}a. This single-valued curve corresponds to the sum of the voltages across all $N$ junctions, $V(I) = \sum v_i(I)$.   For junctions with nearly equal $v_i(I)$, we set  $V(I) = N v(I)$, which allows deducing the average $v(I)$ per junction from the experimentally measured $V(I)$.   The tunneling spectrum $\sigma(v) \equiv {\rm d}I / {\rm d}v(v)$ can then be obtained by a numerical differentiation of the $v(I)$-curve. 

Fig.~\ref{comparison}a shows $\sigma(v)$ for the Bi2201-La$_{0.4}$ sample at several temperatures.  Two sets of peaks are clearly seen at voltages of $\pm v_s = 11.5$ and $\pm v_p = 17$~mV  per tunnel junction.  The inner peaks rapidly smear with temperature and completely fade away close to $T_c$, implying that they are closely related to superconductivity.  The outer ones, on the contrary, change very little and survive even at $T>T_c$, representing a pseudogap.  

Surprisingly, both peaks are equally sharp at low temperatures, and the outer peaks are even higher than the inner ones.  These spectra should be compared with a typical spectrum obtained on Bi2212 \cite{HgBr}, see Fig.~\ref{comparison}b, where
the outer peaks (humps) are very shallow in agreement with many tunneling and ARPES studies \cite{Renner,Zasadzinski,Campuzano}.  The humps and superconducting peaks are separated by "dips" \cite{Zasadzinski}.  In both compounds, the PG peaks persist nearly to room temperature, while the SG peaks disappear at $T_c$ \cite{HgBr,Krasnov}. 

Surface-tunneling and ARPES experiments usually show that the superconducting peaks at $T<T_c$ transform gradually into the pseudo-gap humps at $T > T_c$ \cite{ARPES,Renner}.  Such observations promote the idea that the humps and superconducting peaks have a common origin as in the model of precursor superconductivity \cite{precursor}.  In this model,  the superconductivity pairing actually starts at a temperature much higher than $T_c$,  giving rise to partial depletion of the DOS (pseudogap).  These pre-formed Cooper pairs form a coherent superconducting condensate first at a lower temperature, $T_c$, and the corresponding features of the tunneling spectra sharpen.

In contrast, the temperature evolution of the intrinsic spectra systematically shows that humps are distinct from superconducting peaks at all temperatures implying that the pseudo- and superconducting gaps are independent parameters and most likely have different origins, see Figs \ref{comparison} and \ref{3D} and Refs. \cite{HgBr,Krasnov}.   It should be emphasized that the humps are obviously the leading features of the intrinsic tunneling spectra in the present experiments, with dips just being a result of a superposition of the superconducting peaks and humps at different voltages, contrary to Ref. \cite{Zasadzinski}.

Assuming that the peaks represent different energy gaps in the system, we can obtain numerical values for them, taking $2\Delta_s = v_s$ and $2\Delta_p = v_p$. Next, we take the temperature of the minimum in the $c$-axis resistance-versus-temperature curve, $R_c(T)$, as a  measure of the characteristic pseudogap temperature $T^*$. In the case of Bi2212, $T^*$ coincides with $T^*_\chi$ determined from susceptibility measurements at all doping levels \cite{Watanabe}. In our experiments, $T^*$ both for the Bi2212 and Bi2201 samples roughly matches the temperature when the depression of $\sigma(0)$ smears out and $\sigma(V)$ becomes convex with temperature.  To illustrate this, we plot in Fig.~\ref{PG_T} several intrinsic tunneling spectra of another Bi2201-La$_{0.4}$ sample at temperatures around the minimum in $R_c(T)$.  It is seen that the two rather broad PG peaks decrease in hight with temperature and eventually go under the bell-shaped background right near $T\approx T^*$.  

The ratio $2\Delta_s/k_BT_c \approx 4.2$ is close to the BCS value, while $2\Delta_p/k_BT^* \approx 0.76$ makes no sense.  The ratio $T^*/T_c \sim 8-10$ is much larger than values reported previously for pure Bi2201  and other HTS \cite{Renner}.  Preliminary experiments with overdoped Bi2201-La$_{0.2}$ (corresponding to a hole concentration $p \approx 0.19$ \cite{Ando}) have shown that this ratio is $\sim 3-4$.  For both overdoped and underdoped ($x=0.6$) samples, the quality of the resulting I-V curves was much worse than in Bi2212 for reasons unknown at the moment. They did not clearly show regularly spaced branches and it was difficult to determine the number of junctions. The intrinsic spectra on overdoped and underdoped samples also have strong PG humps, with the latter noticeably more smeared \cite{IJJ_Proc}.

The doping dependencies $T^*(p)$ together with $T_c(p)$ (taken from literature) are shown in Fig.~\ref{phase}. Data points represent the present experiments with $T^*$ defined as above. The doping $p$ was assumed from the nominal La-content according to Ref.~\cite{Ando}.  The best-fit line through the data points for Bi2201-La$_x$ lies above and is steeper than the $T^*(p)$-line for Bi2212.  It intersects the doping axis at $p\approx 0.2$, almost at $p_c=0.19$ indicated as a quantum critical point in several models \cite{Tallon}.  

An important result of this paper is that, despite a difference of a factor of three in the optimal $T_c$s, and different energy scales for the Bi2201-La$_x$- and Bi2212 spectra (see Fig.~\ref{comparison}), the corresponding $T^*(p)$-lines are in close vicinity to each other, see Fig.~\ref{phase}. This fact agrees with the notion that superconductivity and pseudogap are independent phenomena, which has previously been proposed by earlier intrinsic-tunneling measurements on Bi2212 \cite{HgBr,Krasnov}, as well as heat-capacity and nuclear-magnetic-resonance studies \cite{NMR}.

It may be argued that the coincidence simply suggests that the energy scale for the formation of preformed pairs is similar for Bi2201 and Bi2212; the large difference in $T_c$ between the two systems is due to the difference in the strength of the phase fluctuations.  This argumentation does not agree well with another important result that the humps in Bi2201 are much stronger than in Bi2212 at one and the same temperature.  Intense fluctuations in Bi2201 would be expected to smear the spectral features more efficiently.

Although we do not have a proper explanation for why PG humps are much stronger in Bi2201 than in Bi2212, we would like to outline two possible mechanisms.

One mechanism is associated with a saddle-point singularity (usually called van Hove singularity, VHS) in the vicinity of the $(\pi,0)$ point of the Brillouin zone. The existence of the VHS follows from the band structure of a two-dimensional (2D) metal on a square lattice with hopping beyond nearest neighbors: 
$\epsilon_k=-2t(\cos k_x+\cos k_y)+4t'\cos k_x\cos k_y.$  The singularity lies below and close to the Fermi level, depending on $t'\neq 0$, and should result in two symmetric peaks in the tunneling conductance between two such metals. Below $T_c$, there should be two more (superconducting) peaks in the spectra \cite{sis_vhs}. The VHS-scenario of the PG formation was studied theoretically by Markiewitcz \cite{Markiewicz} and Onufrieva \textit{et al} \cite{Onufrieva}.

The intrinsic tunneling spectroscopy is perhaps the most feasible method where these VHS peaks can be detected. Ideal alignment of the single-crystalline electrodes assures that the intrinsic tunneling occurs with conservation of $k_{x,y}$ (specular tunneling). Such a tunneling would take account of the details of the Fermi surface, in particular, VHS. The $\sigma(V)$ should be calculated in terms of the spectral functions $A({\bf k},E)$ \cite{Eschrig} instead of the overall DOS \cite{sis_vhs}.  Then, the intrinsic tunneling conductivity, $\sigma(V)$, can be readily associated with the peculiar ARPES spectra at the $(\pi,0)$ point where sharp peaks and humps are seen even at room temperature \cite{Fink}.

Since an intra-bilayer coupling between individual CuO planes is expected to smear VHS \cite{Markiewicz}, the PG humps should be weaker in the double-layer Bi2212 as compared to the single-layer Bi2201. 2D VHS can also lead to charge- or spin-density waves \cite{Onufrieva,Markiewicz,Gabovich}, which would account for the large increase of $R_c(T<T^*)$, see Fig.~\ref{PG_T}.

Another mechanism for the sharp and strong PG peaks is the long-known resonant tunneling which, in application to HTS, was studied theoretically by Abrikosov \cite{Abrikosov}. 

Indeed, the Bi2212 and Bi2201 single crystals are by themselves just periodic multibarrier tunneling structures that are, by geometry, reminiscent of 2D semiconducting multilayer heterostructures.  An overall transmission coefficient for tunneling through such structures is known to have  several sharp peaks \cite{multilayers}, which should be largely independent of superconductivity in the system making sort of "background" for all other features.  

Even one intrinsic junction consists of several layers. The intermediate double-BiO ones are situated inside the intrinsic tunneling barriers between the CuO$_2$ "electrodes" and therefore can represent wells and resonant centers. The PG signatures in optical and d.c. conductivities can be well modelled assuming that the $c$-axis tunneling involves these layers \cite{Atkinson}.

Although these suggested models are not supported here by any thorough analysis, they draw attention to quite simple mechanisms for experimental observations and invite to a more quantitative theoretical consideration.

In conclusion, we have experimentally found that despite quite different superconducting properties of Bi2201-La$_x$ and Bi2212, their characteristic pseudogap temperatures are nearly the same, suggesting that the superconductivity and pseudogap are independent phenomena in HTS.  Furthermore, the PG "humps" are seen as strong and distinct features of Bi2201 intrinsic spectra, much stronger than previously reported. These new findings require further theoretical understanding. 

\acknowledgments
Financial support from the Swedish Research Council and the Swedish Foundation for Strategic Research (OXIDE program) is greatly appreciated.

%Ref.~\cite{Tallon}, and $T_c(p)$ for Bi2201-La$_x$ - from Ref.~\cite{Ando}.

\end{multicols}

\newpage

\begin{figure}[bth]
\begin{center}
\vspace{0.5cm}
\caption{A low-bias I-V characteristic of a Bi2201-La$_{0.4}$- mesa with $N = 18$ intrinsic tunnel junctions at $T = 4.2$~K. a) The I-V characteristic becomes single valued at high current.  The rectangle denotes the region which is zoomed in the main panel.  b) The voltage measured at 0.3~mA for every counted branch ($n = 1\ldots N$).  The perfectly linear dependence indicates the uniformity of the junctions in the mesa. } 
\label{dIV1}
\end{center}
\end{figure}

\begin{figure}[bth]
\begin{center}
\vspace{0.5cm}
\caption{A comparison of intrinsic tunnelling spectra for a): Bi2201-La$_{0.4}$  and b): Bi2212  at different temperatures.  The voltage scale is per junction.  The spectra were obtained by numerical differentiation of the sum-voltage curves, see Fig.~\ref{dIV1}.   The curves at $T > 4.5$~K are shifted downwards for clarity. } 
\label{comparison}
\end{center}
\end{figure}

%\newpage
\begin{figure}[bth]
\begin{center}
\vspace{0.5cm}
\caption{A false-color  three-dimensional plot of temperature-dependent intrinsic tunnelling spectra of Bi2201-La$_{0.4}$.  Horizontal lines denote the temperatures at which the spectra were taken.  The thick dots mark positions of conductivity maxima.   The temperature is in a logarithmic scale.} 
\label{3D}
\end{center}
\end{figure}

%\newpage
\begin{figure}[bth]
\begin{center}
\vspace{0.5cm}
\caption{The intrinsic tunneling spectra of a Bi2201-La$_{0.4}$ sample at temperatures around the minimum in $R_c(T)$ at $T\approx T^*$ (inset).} 
\label{PG_T}
\end{center}
\end{figure}

\begin{figure}[bth]
\begin{center}
\vspace{0.5cm}
\caption{A pseudogap phase diagram for Bi2212 (dashed line) and Bi2201-La$_x$ (solid lines and dots).  The straight solid line represents the best linear fit through the data points for Bi2201-La$_x$.  $T^*$ and $T_c(p)$-lines for Bi2212 were adopted from Ref.~[14], and $T_c(p)$ for Bi2201-La$_x$ - from Ref.~[9].} 
\label{phase}
\end{center}
\end{figure}

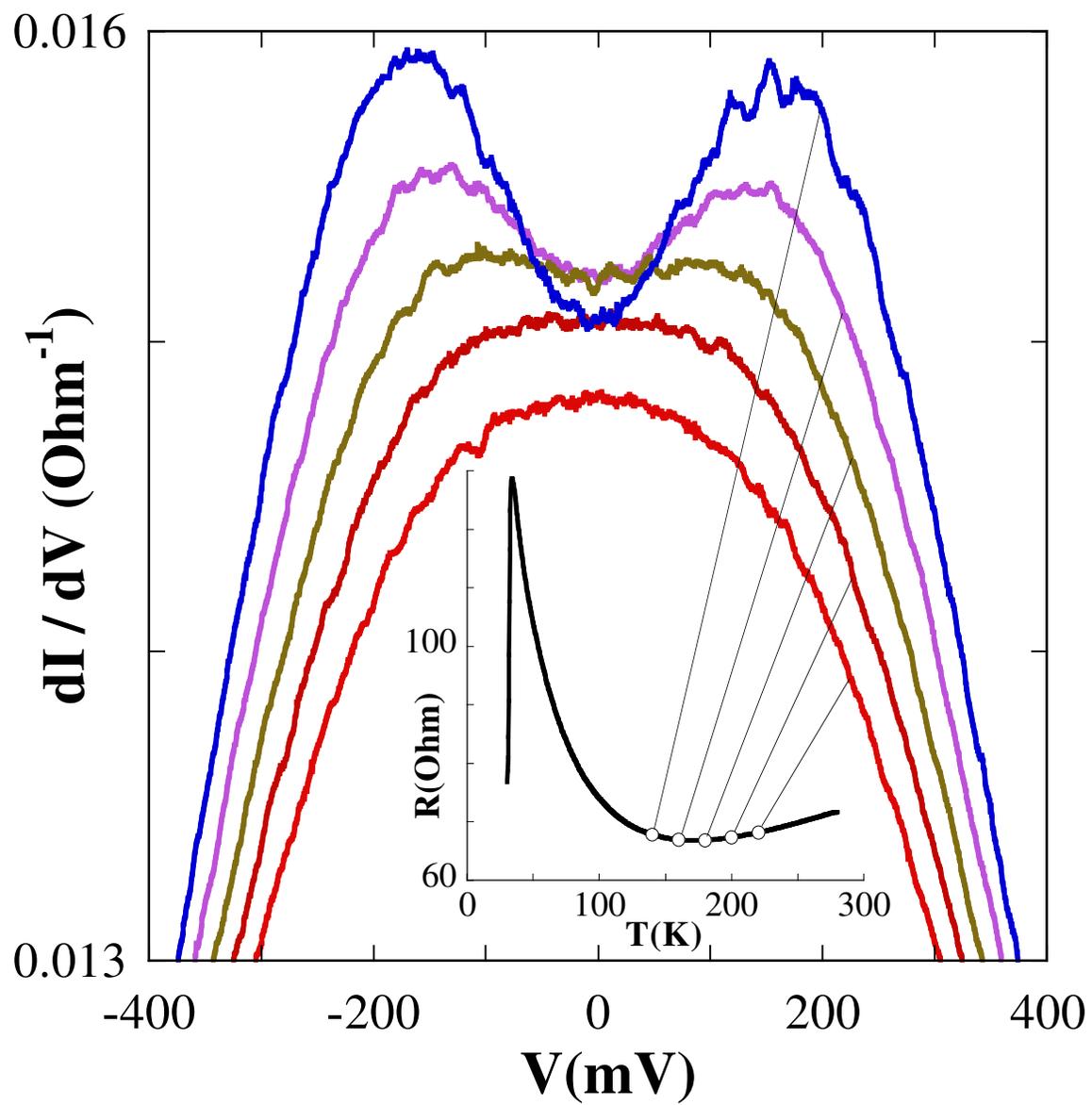
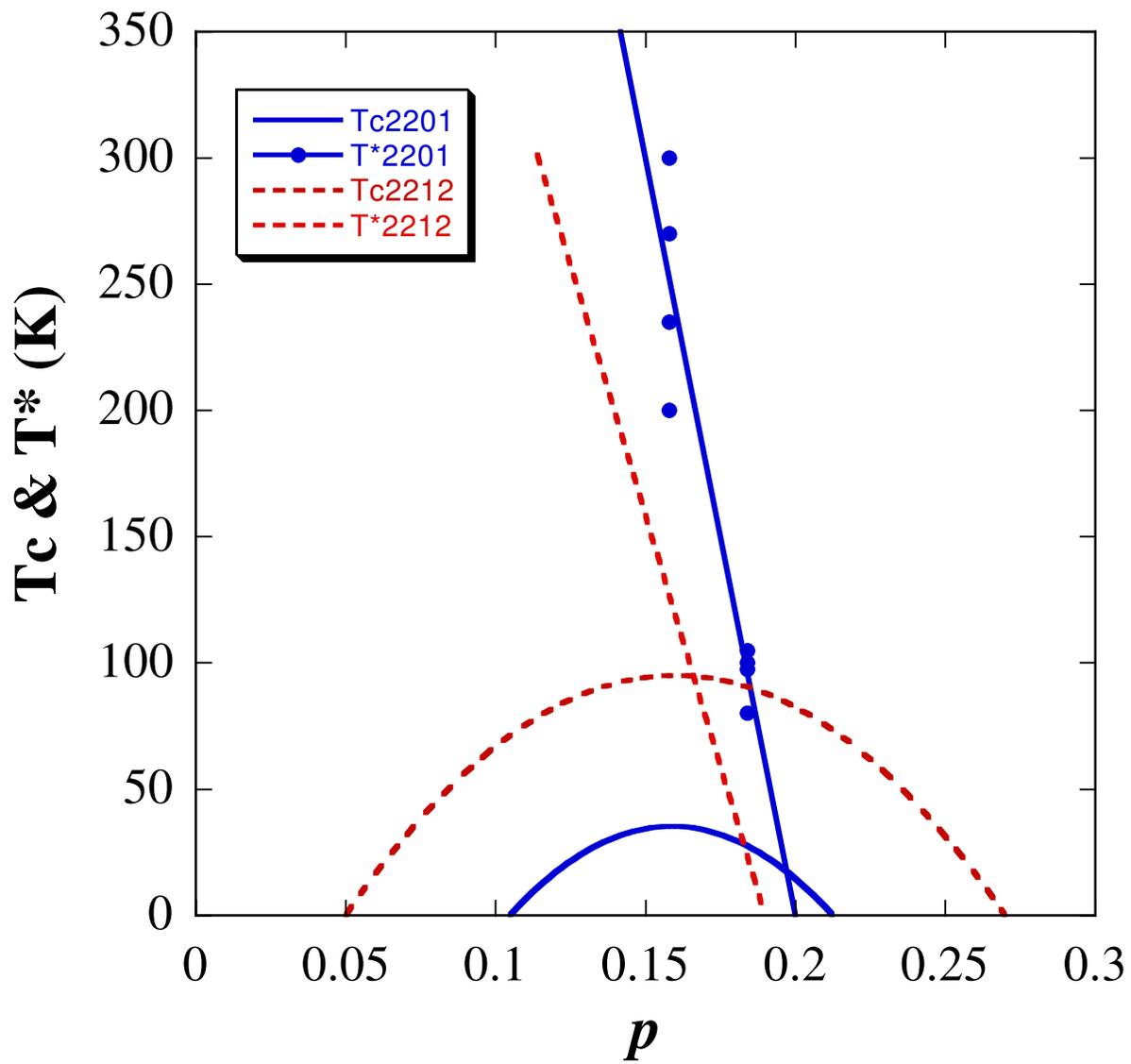
\end{document}